\documentstyle[11pt]{article}

\def\salto{\hspace*{8mm}}
\begin{document}
\def \vsigma{\overrightarrow{\sigma}}
\def \vgamma{\overrightarrow{\gamma}}
\def \vh{\overrightarrow{h}}
\def \msigma{\overline{\sigma^m}}
\def \hsigma{\overline{\sigma^h}} 
\def \Tr{{\rm Tr}}
\def \Re{{\rm Re}}
\def \Im{{\rm Im}}
\def \ds{\overline{d}}
\newtheorem{definition}{Definition}
\newtheorem{lemma}{Lemma}               
\newtheorem{theorem}{Theorem}               
\begin{titlepage}
\begin{center} 
{\bf 
\begin{Large} 
Magnetization bound for classical spin models on graphs\\
\end{Large} } 
\vspace*{1.75cm}
{\large Raffaella Burioni\footnote{E-mail: burioni@pr.infn.it}},
{\large Davide Cassi}\footnote{E-mail: cassi@pr.infn.it}
         {\large and Alessandro Vezzani}\footnote{E-mail: vezzani@pr.infn.it}
\\[.125cm]
          Istituto Nazionale di Fisica della Materia, Unit\`a di Parma\\
          Dipartimento di Fisica, Universit\`a di Parma\\
          and INFN, Gruppo Collegato di Parma,\\
          Viale delle Scienze, 43100 Parma, Italy \\[1cm]
         \end{center}
\vskip.125cm
\begin{abstract} 
In this paper we prove the existence of phase transitions at finite temperature
for $O(n)$  classical ferromagnetic spin models on infrared finite graphs. 
Infrared finite graphs are infinite graphs with  $\lim_{m\to 0^+}{\overline
Tr}(L+m)^{-1}<\infty$, where $L$ is the Laplacian operator  of the graph. The
ferromagnetic couplings are only requested to be uniformly bounded  by two
positive constants. The proof, inspired by the classical result of Fr\"ohlich, 
Simon and Spencer on lattices, is given through a rigorous bound on the average 
magnetization. 
The result holds for $n\ge 1$ and it includes as a particular case the 
Ising model.
\end{abstract}
\vfill
\end{titlepage}

\pagenumbering{arabic}
\setcounter{page}{1}

\section{Introduction}The study of statistical models on graphs requires the
introduction of new techniques  and concepts with respect to the well known
case of lattices. The lack of translational invariance in general implies
inhomogeneity in local variables and makes useless the introduction of such a
powerful tool as Fourier transforms in space. In addition, due to the absence
of a natural definition of dimension, the statement itself of  general results
and theorems is rather problematic. The key point is to find a connection 
between geometry and physical properties. This problem can be studied in the
framework of  algebraic graph theory. However, phase transitions only occur on
infinite graphs and the  algebraic  theory of infinite graphs is a very recent
field of research in mathematics \cite{gerl, woess}. Up to now only
few classical results on lattices have been extended to graphs.
                      
In this paper we deal with the generalization of the proof of existence of spontaneous 
magnetization at finite temperature for ferromagnetic models on 
lattices in $d\ge 3$.
This result was proven in 1976 by Fr\"ohlich, 
Simon and Spencer \cite{FSS} by a general approach based on Gaussian 
domination of correlation functions 
in the infrared regime.                 Such approach, as well as the
formulation of the result itself, is deeply related to the translation
invariance of lattices.  Therefore here we modify the mathematical techniques
and we use an alternative definition of the order parameter. In particular the
concept of  dimension is replaced by the asymptotic behavior of the Laplacian
spectral density at low eigenvalues,  a choice which is meaningful also in 
many other contexts. 
The structure of the paper is as follows: in section 2 we recall the basic
definitions and theorems of graph theory to be used in the proof; in section 3
we give the fundamental known results about statistical models on graphs and
state our theorem on the  magnetization bound; finally in section 4 we present
the proof.    
\section{Some mathematical properties of graphs}
\begin{definition}
A graph G consists of a countable set of vertices (points, sites) 
$V(G)=\{i,j,k,\ldots \}$ and of a set $E(G)$ of unordered pairs of vertices. 
The generic element $(i,j)$ of
$E(G)$ is called a link and its vertices are said to be nearest neighbors
(adjacent). If $V(G)$ is finite, $G$ is called a finite graph and we will
denote by $N$ the number of vertices of $G$. 
\end{definition}
\begin{definition}
A path connecting two vertices $i$ and $j$ is an alternating sequence of 
vertices and links $[i, (i,k), k, (k,h), h \ldots, w, (w,j), j]$.
\end{definition}
\begin{definition}
A graph $G$ is connected if, given any two vertices $i, j \in V(G)$, it exists a
path between $i$ and $j$.
\end{definition}
Here we will deal only with connected graphs.
The topology of $G$ is algebraically described by its adjacency matrix.
\begin{definition}
The adjacency matrix of $G$, $A_{ij}$, is given by:
\begin{equation}
A_{ij}=\left\{
\begin{array}{cl}
1 & {\rm if } \ (i,j)\  \in E(G)\cr
0 & {\rm otherwise}\cr
\end{array}
\right .
\label{defA}
\end{equation}
\end{definition}
Most physical models on $G$ can be defined in terms of the Laplacian 
matrix $L_{ij}$:
\begin{definition}
The Laplacian matrix of $G$, $L_{ij}$ is given b:
\begin{equation}
L_{ij} = z_i \delta_{ij} - A_{ij}
\label{defL}
\end{equation}
where $z_i\equiv\sum_{j\in V(G)} A_{ij}$ is called the coordination 
number of $i$.  
\end{definition}
In the following we will consider graphs with coordination numbers 
bounded from above: $z_i \le z_{max} \in {\bf N} ~~ \forall i \in V(G)$.
If $G$ is infinite, i.e. $V(G)$ is infinite, $L_{ij}$ can be considered as the
representation of the Laplacian operator $L$ acting on $\ell_2(V(G))$. Under
the previous conditions for $G$, it can be shown that $L$ is symmetric, non
negative and bounded \cite{woess}.  
A relevant property of $L$ we will exploit in the
following is the Schwinger-Dyson identity \cite{hhw}. Let us consider a
diagonal bounded operator $\eta_{ij} \equiv \eta_i \delta_{ij}$ with
$ \eta_i \in ${\bf C} such that  
$(L + \eta)^{-1}_{ij}$ exists. Then:
\begin{equation}
\sum_{i \in V(G)} \eta_i (L + \eta)^{-1}_{ij}=1 .
\label{SD}
\end{equation}
This identity follows from the property $\sum_i L_{ij}=0$.

The graph $G$ is naturally provided with an intrinsic metric defined by the
chemical distance on $G$. \begin{definition} The chemical distance $r_{i,j} \in
{\bf N}$ between two vertices $i$ and  $j$ is the number of links in the
shortest path connecting $i$ and $j$.  \end{definition} The intrinsic metric is
the fundamental tool to define the thermodynamical limit  on a infinite graph
$G$, allowing us to introduce the generalized Van Hove spheres.
\begin{definition}
The Van Hove sphere $S_{i,r}$ of center $i$ and radius $r$ is  
the subset of $V(G)$, $S_{i,r}= \{j \in V(G)) ~| r_{ij} \le r\}$.
We will define $N_{i,r}$ as the number of vertices in $S_{i,r}$. 
\end{definition}
The behaviour of $N_{i,r}$ as a function of $r$ characterizes the growth
of $G$. In particular if 
\begin{equation}
A_i r^c \leq N_{i,r} \leq B_i r^c \ 
{\rm where}~ c, \ A_i ~ {\rm and}\ B_i \ {\rm are \ positive \ constants} 
~\forall i.
\label{defc}
\end{equation}
$G$ is said to have polynomial growth and we define the connectivity exponent
$c$  to be the inf of the set of all $c$ satisfying (\ref{defc}).
Here we will consider graphs with polynomial
growth. Indeed this condition is sufficient to guarantee the convergence of the 
restriction $O_{i,r}$ to $S_{i,r}$ of a bounded operator $O$ 
defined on $G$ to $O$ itself for $r\to\infty$.  
\begin{definition}Let $G$ be an infinite graph satisfying all previous
conditions. Given a complex function of the vertices $\phi: V(G) \rightarrow
{\bf C}$ we define the thermodynamic limit average $\overline 
\phi$ of $\phi$ as:
\begin{equation}
\overline{\phi}=\lim_{r\rightarrow\infty}
{1\over \displaystyle N_{i,r}} \sum_{j\in S_{i,r}} \phi_j .
\label{deftd}
\end{equation}
$\overline \phi$ can be easily shown to be independent of $i$.
\end{definition}
In this work we will use also on finite graphs the symbol $\overline{\phi}$ to 
indicate the average over all vertices. We will define the average on a subset
$V_\lambda \subseteq V(G)$:
\begin{equation}
\overline{\chi_{V_{\lambda}} \phi}=\lim_{r\rightarrow\infty}
{1\over \displaystyle N_{i,r}}
\sum_{i\in S_{i,r}}\chi_{V_{\lambda}}(i)\phi_{i}  .
\label{deftd2}
\end{equation}
where $\chi_{V_{\lambda}}(i)$ is the characteristic function of $V_{\lambda}$.
We define the measure $\mu(V_{\lambda})$ of $V_{\lambda}$:
\begin{equation}
\mu(V_{\lambda})= \overline {\chi_{V_\lambda}} .
\label{measure}
\end{equation}
More generally, the thermodynamic limit average of a $k$ variable function
$\Phi: V((G))^k \rightarrow ${\bf C} is defined by:
\begin{equation}
\overline \Phi =\lim_{r\rightarrow\infty}{1\over \displaystyle N_{i,r}} 
\sum_{i_1,\ldots, i_k \in S_{i,r} } \Phi_{i_1,\ldots, i_k} .   
\end{equation}	                                                  
We define the thermodynamic limit average trace of an infinite 
matrix $B_{ij}$ as:
\begin{equation}
\overline{\Tr}B= \overline b 
\label{dtr}
\end{equation}
where $b_i \equiv B_{ii}$, and its restriction to a subset 
$V_\lambda \subseteq V(G)$:
\begin{equation}
\overline{\Tr}_{V_{\lambda}} B \equiv \overline{\chi_{V_{\lambda}} b}
\end{equation}

The large scale topology affecting the critical behavior of statistical models
on graphs  can be characterized by the properties of the trace of $(L+M)^{-1}$,
where  $M_{ij}= m ~\delta_{ij}$
is a real and positive diagonal matrix.  
\begin{definition}
An infinite graph is infrared finite if
\begin{equation}
\lim_{m \rightarrow 0^+} \overline{Tr}(L+M)^{-1} < \infty.  
\label{limtr}
\end{equation}
An infinite graph is infrared infinite if:
\begin{equation}
\lim_{m \rightarrow 0^+} \overline{Tr}(L+M)^{-1} = \infty.  
\label{ilimtr}
\end{equation}
\end{definition}
It can be shown that if we  substitute the constant 
matrix $M$ with $M'_{ij} = m ~b_i ~\delta_{ij}$ where $0<\epsilon < b_i < K $
the infrared behavior for $m\to 0^+$ does not change \cite{debole}. 

In dealing with statistical models the adjacency matrix $A_{ij}$ 
is often generalized 
to the ferromagnetic interaction matrix $J_{ij}$:
\begin{definition}
\begin{equation}
J_{ij} = J_{ji} = \left\{
\begin{array}{cl}
J_{ij} & \ {\rm with\ } 0< \epsilon\leq J_{ij}\leq J\ {\rm  if\ } A_{ij}=1 \cr
0      &  \ {\rm  if\ } A_{ij}=0 \cr 
\end{array}
\right .
\label{defJ}
\end{equation}
In this case the Laplacian generalizes to:
\begin{equation}
L_{ij} = z_i \delta_{ij} - J_{ij}
\label{defgL}
\end{equation}
where $ z_i =\sum_j J_{ij}$. 
\end{definition}
The generalized Laplacian has the same properties of 
the Laplacian associated to $A$: it is symmetric, positive and bounded, it
satisfies  the Schwinger-Dyson identity and finally 
it has the same infrared behaviour \cite{forte}.
 
\section{O(n) classical spin models on graphs}

An important class of  classical statistical spin models on graphs 
is defined by the Hamiltonian:
\begin{equation}
H' = - {1\over 2}\sum_{i,j\in V(G)} J_{ij} 
\vsigma_i \cdot \vsigma_j  -\vh \cdot \sum_{i \in V(G)} \vsigma_i
\label{defH}
\end{equation}
where $J_{ij}$ is a ferromagnetic interaction matrix defined on the graph $G$
satisfying  conditions (\ref{defJ}) and  $\vsigma_j$ are n-dimensional real
unit vectors $\vsigma_i \equiv (\sigma^1_i,\ldots,\sigma^n_i)$  defined on each
vertex $i$ and satisfying the constraints:
\begin{equation}
\vsigma_i^2=1 ~~ \forall i .
\label{const}
\end{equation}
For $n=1$ $H'$ defines the Ising model which is invariant under the discrete 
symmetry group ${\bf Z_2}$, while for $n \geq 2$ $H'$ represents an $O(n)$ 
model with continuous symmetry. 
Finally $\vh \equiv (h,0,\ldots,0)$ is an external magnetic 
field coupled to $\vsigma_i$. In the following we will set $h>0$.

Due to (\ref{const}) $H'$ is equivalent up to an additive constant to:
\begin{equation}
H = {1\over 4}\sum_{ij} J_{ij} (\vsigma_i -\vsigma_j)^2  
-\vh \cdot \sum_i \vsigma_i 
= {1\over 2}\sum_{ij} L_{ij} \vsigma_i \cdot \vsigma_j  
-\vh \cdot \sum_i \vsigma_i
\label{newH}
\end{equation}
where the Laplacian operator defined in (\ref{defgL}) has been introduced.

The Boltzmann measure $\mu_{\beta,h} (\vsigma)$ is given by:
\begin{equation}
d \mu_{\beta,h}(\vsigma) \equiv \prod_i d\vsigma_i e^{-\beta H}~
\label{boltz}
\end{equation}
where $\beta=1/k_B~T$, with $\vsigma_i$ satisfying the constraints
(\ref{const}).
The order parameter of the model
is the average magnetization in the $\vh$ direction:
\begin{equation}
M (\beta,h) =  {\displaystyle \int   d \mu_{\beta,h}(\vsigma) ~ \overline
{\sigma^1}    \over \displaystyle \int d \mu_{\beta,h}(\vsigma)} .
\label{defm}
\end{equation} On lattices, in the thermodynamical limit for $n\ge 2$ it is
known that the limit  for $h\rightarrow 0$ of the order parameter vanishes for
$T > 0$ in the infrared  infinite case (i.e. if the Euclidean dimension $d$ is
2 or 1) \cite{mw}. In the infrared  finite  case ($d \ge 3$) for low enough
temperatures the limit is positive \cite{FSS}, implying the existence of
phase transitions. On graphs we have
$\lim_{h\rightarrow 0}M(\beta,h)=0$ for every  $\beta$, if for 
each positive measure subset $V_{\lambda} \subseteq V(G)$ we have \cite{mwg}:
\begin{equation}
\lim_{b\rightarrow 0^+}\overline{\Tr}_{V_{\lambda}}(L+b)^{-1}= \infty~.
\label{mermin}
\end{equation}
For Ising models on lattices the limit of the order parameter 
vanishes for all temperatures only in one dimension. There are no analogous
results for generic graphs.

In this work we will show that, for the classical spin models (\ref{defH}) on 
infrared finite graphs, below  a certain 
temperature $M (\beta,h)$ satisfies the bound:
\begin{equation} 
\lim_{h\rightarrow 0}M (\beta,h) > k >0.   
\label{bound}
\end{equation}
This result for $O(n)$ models is not the exact inversion of \cite{mwg}. Indeed  
one should prove that if there is a positive measure subset 
$V_{\lambda}$, for which the limit in (\ref{mermin}) is finite, then the 
bound (\ref{bound}) is satisfied. Here the proof is given only for the case 
where the limit in (\ref{mermin}) is finite for $V_{\lambda} \equiv G$. 
A complete inversion of \cite{mwg} will be given in a following paper
\cite{invmwg}.  We remark that this is the first result for the Ising model on
a generic graph.
                                                                      
\section{The magnetization bound}       
In this section we will prove the magnetization bound (\ref{bound}) for the 
classical spin model (\ref{defH}) according to the following strategy:

(A) first of all we will prove some useful inequalities for the Laplacian on 
finite graphs;
 
(B) then we will introduce for the constraint (\ref{const}) an integral 
representation allowing us to perform a Gaussian integral with respect to
the spin variables $\vsigma_i$;

(C) we will take the thermodynamic limit of the inequalities (A) and in 
the expressions for $M$ and for the global constraint;

(D) exploiting a saddle point technique for large $\beta$, we will apply 
the inequalities in (C) to $M$ getting the bound
(\ref{bound}), for infrared finite graphs.

0.3truecm
As for the point (A):
\begin{lemma} 
The following inequalities hold for finite graphs, 
with $h_{ij} \equiv h ~\delta_{ij}$ and $\alpha \equiv \alpha_i ~\delta_{ij}$ 
with $\alpha_i \in {\bf R}~\forall i$:

\begin{equation}
1 \geq \Re\left( {h\over N} \sum_{ij}(L+h+i\alpha)_{ij}^{-1} \right)\geq
\Re\left( {h^2\over N} \sum_{ij}(L+h+i\alpha)_{ij}^{-2}\right) .
\label{ineq1}
\end{equation}

\begin{equation}
0 \leq {1\over N}\sum_{ikj}\alpha_i(L+h-i\alpha)_{ik}^{-1}(L+h+i\alpha)_{kj}^{-1}\alpha_j
\leq\Re\left({1\over N h}\sum_{ij}\alpha_i(L+h+i\alpha)_{ij}^{-1}\alpha_j\right)
\label{ineq4}
\end{equation}

\begin{equation}
0 \leq {1\over N}\Re\Tr_{V_{\lambda}}\left(L+h+i\alpha)^{-1}\right) \leq
{1\over N}\Re\Tr_{V_{\lambda}}\left((L+h)^{-1}\right) \leq
{1\over N}\Re\Tr\left((L+h)^{-1}\right)
\label{ineq2}
\end{equation}

\begin{equation}
{1\over N}| \Im \Tr_{V_{\lambda}} (L+h+i\alpha)^{-1}| 
\leq {1\over 2N} \Tr _{V_{\lambda}}(L+h)^{-1} \leq {1\over2 N} \Tr (L+h)^{-1}
\label{ineq3}
\end{equation}
where $V_{\lambda}$ is a generic subset $V(G)$ and
\begin{equation}
\Tr_{V_{\lambda}} B \equiv \sum_i B_{ii}~\chi_{V_\lambda}(i)~,
\end{equation}
\begin{equation}
{1\over N} \sum_i \alpha_i^2  \leq f(\alpha)  \left( {1\over N}\sum_{ikj} 
\alpha_i(L+h-i\alpha)_{ik}^{-1}(L+h+i\alpha)_{kj}^{-1}\alpha_j\right)^{1/2}
\label{ineq5} 
\end{equation}
where $f(\alpha) \leq 1/N \sum_i (\alpha_i^2(l_{max}+h)^2+\alpha_i^4)$
and $l_{max}$ is the maximum eigenvalue of $L$.
\end{lemma}
The proof of these inequalities will be given at the end of this section. $\Box$
\salto 

As for (B), we prove the following:
\begin{lemma}
On a finite graph the expression of the average magnetization (\ref{defm})
and of the constraint:
\begin{equation}
1 = {\displaystyle \int \overline{\vsigma^2} 
  d\mu_{\beta,h}(\vsigma) 
\over \displaystyle \int d\mu_{\beta,h}(\vsigma)}   
\label{glconst}
\end{equation}
can be written in the following way:

\begin{equation}                                    
M(\beta,h) ={1\over Z}{\int d \mu_{\beta,h}(\alpha) 
\ {h\over N} \sum_{kj}(L+h+i\alpha)^{-1}_{kj}}
\label{defm1}
\end{equation}

\begin{equation}
1 = {1\over Z} \int d \mu_{\beta,h}(\alpha) \left( 
{n \over \beta N} {\Tr}(L+h+i\alpha)^{-1} +
{h^2\over N} \sum_{ij}(L+h+i\alpha)^{-2}_{ij} \right)
\label{glconst1}
\end{equation}
where:
\begin{equation}
d \mu_{\beta,h}(\alpha)\equiv 
\prod_{i\in G} d \alpha_i e^{i S_{\beta,h}}
\label{dmu}
\end{equation}
$$ i S_{\beta,h}\equiv
-{n\over 2} \Tr\left(\ln (L+h+i\alpha)\right) 
+{\beta\over 2}
\left(i\sum_i\alpha_i+h^2\sum_{ij}(L+h+i\alpha)^{-1}_{ij}\right)$$ 
{\rm and}
$$Z\equiv \int d \mu_{\beta,h}(\alpha_i)$$
\end{lemma}
Let us write the constraints (\ref{const}) using the 
complex integral representation of the delta function:
$$\delta(x^2)=\int d\alpha \exp(-i\alpha x^2/2-\epsilon x^2/2)$$
where $\epsilon$ is a real arbitrary constant. We will put 
$\epsilon=h\beta$. 
Substituting the expression for $\delta$ in (\ref{defm}) and (\ref{glconst}) 
we obtain:
$$M(\beta,h)= {\displaystyle \int d \vsigma  \ d\alpha \ \hsigma 
\exp\left(-\beta({1 \over 2} \sum_{ij}L_{ij}{\vsigma_i}\cdot{\vsigma_j} 
-h \sum_i \sigma_i^1) 
-\sum_i(i {\alpha_i \over 2} ({\vsigma_i}^2-1)
-{h \beta \over 2} {\vsigma_i^2})\right)
\over \displaystyle \int d\vsigma \ d\alpha
\exp\left(-\beta({1 \over 2} \sum_{ij}L_{ij}{\vsigma_i}\cdot{\vsigma_j} 
-h \sum_i \sigma_i^1) 
-\sum_i(i {\alpha_i \over 2} ({\vsigma_i}^2-1)
-{h \beta \over 2} {\vsigma_i^2})\right)}$$
and
$$1 = {\displaystyle \int d \vsigma  \ d\alpha \ \overline{\vsigma^2} 
\exp\left(-\beta({1 \over 2} \sum_{ij}L_{ij}{\vsigma_i}\cdot{\vsigma_j} 
-h \sum_i \sigma_i^1) 
-\sum_i(i {\alpha_i \over 2} ({\vsigma_i}^2-1)-
{h \beta \over 2} {\vsigma_i^2})\right)
\over \displaystyle \int d\vsigma \ d\alpha
\exp\left(-\beta({1 \over 2} \sum_{ij}L_{ij}{\vsigma_i}\cdot{\vsigma_j} 
-h \sum_i \sigma_i^1) 
-\sum_i(i {\alpha_i \over 2} ({\vsigma_i}^2-1)-
{h \beta \over 2} {\vsigma_i^2})\right)}$$
where $d\alpha \equiv \prod_i d\alpha_i$ and 
$d\vsigma \equiv \prod_i d\vsigma_i$.
Substituting in both integral $\alpha_i$ with $\beta\alpha_i$ and 
$\vsigma_i$ with $\vsigma_i/\sqrt{\beta}$, we  can perform the Gaussian 
integral on the variables $\vsigma_i$, obtaining in this way (\ref{defm1})
and (\ref{glconst1}). $\Box$

\salto
 
Now we consider an infinite graph $G$ and using the thermodynamic limit, we 
will extend to $G$ the Lemmas (1) and (2). This is point (C) in our plan.

\begin{lemma} 
For an infinite graph the inequalities (\ref{ineq1}), (\ref{ineq4}), 
(\ref{ineq2}), (\ref{ineq3}), (\ref{ineq5}) become:

\begin{equation}
1 \geq \Re\left( h \overline{(L+h+i\alpha)^{-1}} \right)\geq
\Re\left( h^2 \overline {(L+h+i\alpha)^{-2}}\right) .
\label{ineq1'}
\end{equation}

\begin{equation}
0 \leq \overline{\alpha(L+h-i\alpha)^{-1}(L+h+i\alpha)^{-1}\alpha}
\leq\Re\left(\overline{\alpha(L+h+i\alpha)^{-1}\alpha}\right)
\label{ineq4'}
\end{equation}

\begin{equation}
0 \leq \Re\left(\overline{\Tr_{V_{\lambda}}}(L+h+i\alpha)^{-1}\right) \leq
\Re\left(\overline{\Tr_{V_{\lambda}}}(L+h)^{-1}\right)\leq
\Re\left(\overline{\Tr}(L+h)^{-1}\right)
\label{ineq2'}
\end{equation}

\begin{equation}
| \Im \overline{\Tr_{V_{\lambda}}} (L+h+i\alpha)^{-1}| \leq 
{1\over 2}~ \overline{\Tr_{V_{\lambda}}} (L+h)^{-1} \leq
{1\over 2}~ \overline{\Tr} (L+h)^{-1} 
\label{ineq3'}
\end{equation}
here $V_{\lambda}$ is a generic positive measure subset of G.
\begin{equation}
\overline{\alpha^2}  \leq f(\alpha)  \left( \overline{ 
\alpha(L+h-i\alpha)^{-1}(L+h+i\alpha)^{-1}\alpha}\right)^{1/2}
\label{ineq5'} 
\end{equation}
where $f(\alpha) \leq \overline{(\alpha^2(l_{max}+h)^2+\alpha^4)}$.
\end{lemma}
These inequalities are easily obtained by applying the inequalities
(\ref{ineq1'}-\ref{ineq5'}) to the restriction to the Van Hove spheres
$S_{i,r}$ of $L$, $\alpha$ and $h$ and  taking the thermodynamical limit by
letting $r\to\infty$. $\Box$

Now we will prove the main theorem showing the existence of
the bound for the magnetization.

\begin{theorem}
In an infinite and infrared finite graph $G$, $\forall \epsilon >0 \ \exists 
\bar{\beta} >0$ such that $\forall \beta \geq \bar{\beta}$:
\begin{equation}
\lim_{h\rightarrow 0}M(\beta,h)\geq 1 - \epsilon
\label{bound1}
\end{equation}
\end{theorem}
We consider, in the thermodynamic limit, the expressions (\ref{defm1}) and 
(\ref{glconst1}) for $M(\beta,h)$ and for the global constraint, with the 
measure given by (\ref{dmu}). At the end of this section we will prove the 
following lemma:

\begin{lemma}
Consider a positive measure subset $V_{\lambda}$ of $V(G)$.
If $G$ is infrared finite, the quantity
\begin{equation}
\overline {\chi_{V_{\lambda}} S'_{\beta,h}}
\label{sadle1}
\end{equation}
where $S'_{\beta,h}(i)= {\partial\over{\partial \alpha_i}} S_{\beta,h}$ is
bounded for all $V_{\lambda}$ when $\beta\rightarrow\infty$ only if  the
$\alpha_i$ satisfy the condition: 
\begin{equation}        
\overline{\alpha^2}=0 .   
\label{sadle2}
\end{equation}
This condition is true also in the limit $h\rightarrow 0$.  $\Box$
\end{lemma}
We will call the set of values $\alpha_i \equiv \{\alpha_1, \alpha_2, \ldots\}$ 
satisfying (\ref{sadle2}) the stationary point of the statistical weight 
$d \mu_{\beta,h}(\alpha)$.
Now we separate the space of integration variables $\alpha_i$ into two 
subsets $\Gamma_{\beta,h}(\alpha)$ and its complement 
$\overline{\Gamma_{\beta,h}(\alpha)}$ such that $\Gamma_{\beta,h}(\alpha)$
is the region around the stationary point where the real part of the measure
$d \mu_{\beta,h}(\alpha)$ is positive. We define: 
\begin{eqnarray}
\epsilon(\beta,h) & \equiv & {1\over Z} \int_{\overline{\Gamma_{\beta,h}}}
d\Re(\mu_{\beta,h}(\alpha)) \overline{(L+h+i\alpha)^{-1}}  
+ {1\over Z} \int d\Im(\mu_{\beta,h}(\alpha)) 
\overline{(L+h+i\alpha)^{-1}}  \nonumber\\
\epsilon'(\beta,h) & \equiv & {1\over Z} \int_{\overline{\Gamma_{\beta,h}}}
d\Re(\mu_{\beta,h}(\alpha)) \left( {n \over \beta}\overline{\Tr}(L+h+i\alpha)^{-1}  
+ h^2 \overline{(L+h+i\alpha)^{-2}} \right) \label{epsilon}\\
& \ & + {1\over Z} \int d\Im(\mu_{\beta,h}(\alpha)) 
\left({n \over \beta}\overline{\Tr}(L+h+i\alpha)^{-1} +
h^2 \overline{(L+h+i\alpha)^{-2}} \right) \nonumber
\end{eqnarray}
Lemma 4 and the saddle-point theorem imply that:
\begin{equation}
\forall \ \epsilon >0 \ \exists\beta'\in {\bf R}~\ {\rm such ~that}\ 
\forall \ \beta \geq \beta':
\lim_{h\rightarrow 0}|\epsilon(\beta,h)|\leq\epsilon/3 \ {\rm and}   
\ \lim_{h\rightarrow 0}|\epsilon'(\beta,h)|\leq\epsilon/3 .   
\label{beta'}
\end{equation}
Now for $M(\beta,h)$ we have:
\begin{eqnarray}
M(\beta,h) & = & \epsilon(\beta,h) + {1\over Z} \int_{\Gamma_{\beta,h}(\alpha)}
        d\Re(\mu_{\beta,h}(\alpha)) 
        \overline{(L+h+i\alpha)^{-1}} \label{M1}\\
  & = & \epsilon(\beta,h) + {1\over Z} \int_{\Gamma_{\beta,h}(\alpha)}
        d\Re(\mu_{\beta,h}(\alpha)) 
        \ \Re\left( h \overline{(L+h+i\alpha)^{-1}} \right)
        \label{M2}
\end{eqnarray}
The imaginary part  of the second term in (\ref{M1}) does not provide any 
contribution to the integral because $M(\beta,h)$ is a real quantity. In 
analogous way for the constraint (\ref{glconst1}) we get:
\begin{equation}
1= \epsilon'(\beta,h) + {1\over Z} \int_{\Gamma_{\beta,h}(\alpha)}
d\Re(\mu_{\beta,h}(\alpha)) 
\Re \left( {n \over \beta} \overline{\Tr}(L+h+i\alpha)^{-1} 
+ h^2 \overline{(L+h+i\alpha)^{-2}} \right)
\label{gl1}
\end{equation}
Now in (\ref{M2}) we deal with a real quantity averaged with respect to 
a positive measure and we can use the inequalities of Lemma 3. In particular
applying (\ref{ineq1'}) in (\ref{M2}) we get:
$$ M (\beta,h) \geq \epsilon(\beta,h) + {1\over Z} \int_{\Gamma_{\beta,h}
(\alpha)}
d\Re(\mu_{\beta,h}(\alpha))
\ \Re\left( h^2 \overline{(L+h+i\alpha)^{-2}} \right)$$
With this inequality and the expression for the global constraint 
(\ref{gl1}) we get:
\begin{equation}
M (\beta,h)\geq 1 - \epsilon'(\beta,h) + \epsilon(\beta,h)- 
\int_{\Gamma_{\beta,h}(\alpha)} d\alpha \ \Re(\mu_{\beta,h}(\alpha)) 
{n \over \beta} \Re\left(\overline{\Tr}(L+h+i\alpha)^{-1}\right).      
\label{M3}
\end{equation}
Now we take the limit $h\rightarrow 0$ and we study the terms of this 
inequalities for large $\beta$. For $\epsilon(\beta,h)$ and 
$\epsilon'(\beta,h)$ the behaviour is given by (\ref{beta'}).
For the integral term in (\ref{M3}) we can exploit the inequality  
(\ref{ineq2'}) getting:
$$ 0 \leq
\int_{\Gamma_{\beta,h}(\alpha)} d\alpha \ \Re(\mu_{\beta,h}(\alpha)) 
{n \over \beta} \Re\left(\overline{\Tr}(L+h+i\alpha)^{-1}\right) \leq      
{n \over \beta}\overline{\Tr}(L+h)^{-1}
\int_{\Gamma_{\beta,h}(\alpha)} d\alpha \ \Re(\mu_{\beta,h}(\alpha)) .$$      
The saddle-point theorem implies that:
\begin{equation}
\exists  \beta''\in{\bf R} ~\ {\rm such ~that}\ \forall \beta \geq \beta'':
\lim_{h\rightarrow 0} {1 \over Z}
\int_{\Gamma_{\beta,h}(\alpha)} d\Re(\mu_{\beta,h}(\alpha))
\geq 1/2.
\label{beta''}
\end{equation}
If for each $\epsilon$ we choose
\begin{equation}
\beta \geq \beta''~~{\rm and}~~\beta \geq {3 \over \epsilon} 
2n \lim_{h\rightarrow 0}\overline{\Tr}(L+h)^{-1}=\beta'''
\label{beta'''}
\end{equation}
we get:
\begin{equation}
\lim_{h\rightarrow 0}{n\over \beta Z}
\int_{\Gamma_{\beta,h}(\alpha)} d\Re(\mu_{\beta,h}(\alpha)) 
\Re\left(\overline{\Tr}(L+h+i\alpha)^{-1}\right)\leq \epsilon/3.      
\label{epsilon'}
\end{equation}
Finally fixed $0\leq\epsilon\leq1$, if we choose 
$\bar{\beta}=\max(\beta',\beta'',\beta''')$, we have that 
$\forall\beta>\bar{\beta}$ $M(\beta,h)$ satisfies condition (\ref{bound1}).
Note that from (\ref{beta'''}) we have that $\beta'''$ and then $\bar{\beta}$
are bounded only for infrared finite graphs.
$\Box$
 
\subsection{Inequalities}

Now we will prove the inequalities (\ref{ineq1}), (\ref{ineq4}), 
(\ref{ineq2}), (\ref{ineq3}) and (\ref{ineq5}) for finite graphs.
On a finite graph the set of all real functions of the vertices $\phi_i$ has a
vector space structure. Let us define $\langle \phi| = (\phi_1,\ldots,\phi_N)$
and $|\phi \rangle = (\phi_1,\ldots,\phi_N)^t$. The scalar product $\langle
\phi | \psi \rangle$ is defined by:
\begin{equation}
\langle \phi | \psi \rangle = \sum_i \phi_i \psi_i.
\end{equation}
The matrices $L$, $A$, $h$ and $\alpha$  are operators on this space. The matrix
$L$ is diagonalizable by a real transformation, and its eigenvalues $l$ satisfy
$0\leq l\leq l_{max}$. Let us prove the following:

\begin{lemma}
On a finite graph, it is possible to introduce a real operator $B$ defined by:
\begin{equation}
B^t(L+h)B=I\ \ \ \ B^t\alpha B=c
\label{B}
\end{equation}
where $c$ is a real diagonal operator and $I$ is the identity operator. 
Furthermore $B$ satisfies the properties:
\begin{equation}
(L+h+i\alpha)^{-1}=B(1+ic)^{-1}B^t\ \ \ \ \ (L+h)^{-1}=BB^t
\label{propB}
\end{equation}

\begin{equation}
\|B^tB\|={1\over h} 
\label{normB}
\end{equation}
where
$$\|B^tB\|=
\sup_{\phi}{\langle\phi |B^tB |\phi\rangle \over \langle\phi| \phi\rangle}
$$
\end{lemma}
The existence of $B$ is proved in \cite{gant} where it is also shown that is
real. Properties (\ref{propB}) can be immediately obtained by (\ref{B}).
Therefore it is easy to get the exact expression for $B$:
\begin{equation}
B=TAT'
\end{equation}
where $T$ is the orthogonal transformation that diagonalize $L$;
$A$ is the matrix $A_{km}=(1/\sqrt{l_k+h}) \delta_{km}$, where $l_k$ is the 
eigenvalue of $L$ relative to the eigenvector $k$; finally $T'$ is the 
orthogonal operator that diagonalize the symmetric matrix $AT^t\alpha TA$.
$B$ is not an orthogonal transformation but we compute its norm:
$$\|B^tB\|=
\sup_{\phi}{\langle\phi| B^tB |\phi\rangle \over \langle\phi |\phi\rangle}=
\sup_{\phi}{\langle\phi |A^2 |\phi\rangle \over \langle\phi |\phi\rangle}
={1\over h}$$
This proves property (\ref{normB}). $\Box$

\salto

{\it Proof of Lemma 1}
Let $\langle\phi|$ be a generic vector of the space of the functions of 
vertices. We have:
\begin{eqnarray}
\Re \langle \phi|h^2 (L+h+i\alpha)^{-2}|\phi\rangle
& = & \Re h^2 \langle \phi|B(1+ic)^{-1}B^tB(1+ic)^{-1}B^t|\phi\rangle\nonumber\\
& = &  h^2 \langle \phi|B(\Re(1+ic)^{-1})B^tB(\Re(1+ic)^{-1})B^t|\phi\rangle\nonumber\\
& \ & - h^2 \langle \phi|B(\Im(1+ic)^{-1})B^tB(\Im(1+ic)^{-1})B^t|\phi\rangle\nonumber\\
& \leq & h^2 \langle \phi|B(\Re (1+ic)^{-1})B^tB(\Re(1+ic)^{-1})B^t|\phi\rangle\nonumber\\
& \leq & h \langle \phi|B(\Re (1+ic)^{-1})(\Re(1+ic)^{-1})B^t|\phi\rangle\nonumber\\
& \leq & h \langle \phi|B(1+c^2)^{-2}B^t|\phi\rangle\nonumber\\ 
& \leq & h \langle \phi|B(1+c^2)^{-1}B^t|\phi\rangle\nonumber
\end{eqnarray}
where we used properties (\ref{propB}) and (\ref{normB}). Then we 
have:
\begin{eqnarray}
\Re \langle \phi|h (L+h+i\alpha)^{-1}|\phi\rangle
& = &  h \langle \phi|B(\Re (1+ic)^{-1})B^t|\phi\rangle\nonumber\\
& = &  h \langle \phi|B(1+c^2)^{-1}B^t|\phi\rangle\nonumber\\   
& \leq &  h \langle \phi|BB^t|\phi\rangle \leq 
\langle \phi|\phi\rangle\nonumber 
\end{eqnarray}
so we get:
\begin{equation}
\Re \langle \phi|h^2 (L+h+i\alpha)^{-2}|\phi\rangle\leq               
\Re \langle \phi|h (L+h+i\alpha)^{-1}|\phi\rangle \leq \langle \phi|\phi\rangle .
\label{ap2a}
\end{equation}
Now choosing in (\ref{ap2a}) $\langle\phi|$ to be the constant vector
$\alpha_i=1$, we prove (\ref{ineq1}).

To  prove the inequality (\ref{ineq4}) we consider the following expression:
\begin{eqnarray}
0 \leq \langle \phi|(L+h-i\alpha)^{-1}(L+h+i\alpha)^{-1}|\phi\rangle
& = &   \langle \phi|B(1-ic)^{-1}B^tB(1+ic)^{-1}B^t|\phi\rangle\nonumber\\
& \leq & {1\over h} \langle \phi|B(1-ic)^{-1}(1+ic)^{-1}B^t|\phi\rangle\nonumber\\
& \leq & {1\over h} \langle \phi|B(1+c^2)^{-1}B^t|\phi\rangle\nonumber\\ 
& \leq & {1\over h} \Re \langle \phi|B(1+ic)^{-1}B^t|\phi\rangle\nonumber\\
& \leq & {1\over h} \Re \langle \phi|(L+h+i\alpha)^{-1}|\phi\rangle .
\label{ap2b}
\end{eqnarray}
where we used (\ref{propB}) and (\ref{normB}). 
Putting $\phi_i =\alpha_i$ in
(\ref{ap2b}), we get (\ref{ineq4}).

To get inequalities (\ref{ineq2})  and (\ref{ineq3}) we introduce
the base $\langle k|$, $k=1,\ldots,N$ in the vector space of 
the functions of the sites given by:
$$\langle k|=\left\{
\begin{array}{cl}
k_i=1 & {\rm if} \ i=k \\ 
0 & {\rm in\ all\ others\ vertices.}\\
\end{array}
\right .$$
We have:
\begin{equation}
(L+h+\alpha)^{-1}_{kk} = \langle k|(L+h+\alpha)^{-1}|k\rangle=
\langle k|B(1+ic)^{-1}B^t |k\rangle 
\label{lreim}
\end{equation}
For the real part of (\ref{lreim}), with properties (\ref{propB}) we get:
$$0\leq \Re (L+h+\alpha)^{-1}_{kk} = \langle k|B(1+c^2)^{-1}B^t |k\rangle\leq
\langle k|BB^t |k\rangle=(L+h)^{-1}_{kk}$$
If we sum over all $k\in V_{\lambda}$ we get:
$$0\leq\Re \Tr_{V_{\lambda}} (L+h+\alpha)^{-1}\leq 
\Tr_{V_{\lambda}}(L+h)^{-1} \leq \Tr (L+h)^{-1}$$
This proves (\ref{ineq2}). Now for the imaginary part of (\ref{lreim}) we 
have:
$$ |\Im (L+h+\alpha)^{-1}_{kk}| = |\langle k|Bc(1+c^2)^{-1}B^t |k\rangle|\leq
\langle k|B|c|(1+c^2)^{-1}B^t |k\rangle\leq {1\over 2}
\langle k|BB^t |k\rangle={1\over 2} (L+h)^{-1}_{kk}$$
If we sum over all $k\in V_{\lambda}$ we get: 
$$|\Im \Tr_{V_{\lambda}} (L+h+\alpha)^{-1}|\leq 
\sum_{i\in V_{\lambda}} |\Im (L+h+\alpha)^{-1}_{kk}|~
{1\over 2} ~\Tr_{V_{\lambda}}(L+h)^{-1} \leq {1\over 2} \Tr (L+h)^{-1}$$
In this way we proved inequality (\ref{ineq3}). 

For (\ref{ineq5}) we have:
\begin{eqnarray}
\sum_i \alpha_i^2  & = & \langle\alpha|\alpha\rangle
= \langle\alpha|(L+h+i\alpha)(L+h+i\alpha)^{-1}|\alpha\rangle \leq \nonumber\\ 
& \leq &  \langle\alpha|(L+h+i\alpha)(L+h-i\alpha)|\alpha\rangle^{1/2}
\langle\alpha|(L+h-i\alpha)^{-1}(L+h+i\alpha)^{-1}|\alpha\rangle^{1/2} \nonumber\\
& \leq & f(\alpha)
\left(\sum_{ikj}
\alpha_i(L+h-i\alpha)_{ik}^{-1}(L+h+i\alpha)_{kj}^{-1}\alpha_j\right)^{1/2}=0
\nonumber
\end{eqnarray}
where $f(\alpha)=\langle\alpha|(L+h)^2+\alpha^2|\alpha\rangle^{1/2}
\leq [\sum_i (\alpha_i^2(l_{max}+h^2+\alpha_i^4)]^{1/2}$ and this 
completes the proof of Lemma 1. $\Box$

\subsection{The saddle-point condition}

{\it Proof of Lemma 4.} Let us compute the quantity (\ref{sadle1}). We have:
$$ \overline {\chi_{V_{\lambda}} S'_{\beta,h}}
 = -{n\over 2}~~ \overline{\Tr}_{V_{\lambda}} (L+h+i\alpha)^{-1} 
+{ \beta\over 2} \left(\mu(V_{\lambda}) - 
h^2 \overline{(L+h+i\alpha)^{-1}\chi(V_\lambda)
(L+h+i\alpha)^{-1}}\right)$$
where $\chi(V_\lambda)_{ij}=\chi_{V_\lambda}~\delta_{ij}$.
The first term in this expression is always 
bounded, also when $h\rightarrow 0$ for infrared finite graphs; 
inequalities (\ref{ineq2}) and (\ref{ineq3}) can be used for the proof. 
Then for $\beta \rightarrow \infty$
(\ref{sadle1}) is bounded only when the condition:
\begin{equation}
\mu(V_{\lambda}) = 
h^2 \overline{(L+h+i\alpha)^{-1}\chi(V_\lambda)
(L+h+i\alpha)^{-1}}
\label{sadle'}
\end{equation}
is satisfied. Let us consider the particular case where we put in (\ref{sadle'})
$V_{\lambda}=G$. We get
\begin{equation}
1= 
h^2 \overline{(L+h+i\alpha)^{-2}}.
\label{sadle3}
\end{equation}
 Let us show that (\ref{sadle3}) is satisfied only when (\ref{sadle2}) holds.
First of all we have that (\ref{sadle3}) implies:
\begin{equation}
\Re\left(h \overline{(L+h+i\alpha)^{-1}}\right)=1 .
\label{sadle4}
\end{equation}
Indeed with the inequality (\ref{ineq1'}) we get:
$$ 1=\Re\left( h^2 \overline {(L+h+i\alpha)^{-2}}\right)
\leq \Re\left(h \overline{(L+h+i\alpha)^{-1}}\right)
\leq 1 .$$
Now using the Schwinger-Dyson identity (\ref{SD}) in (\ref{sadle4}) we have:
$$1 = \Re\left(h \overline{(L+h+i\alpha)^{-1}}\right)
= 1 + \Re\left(\overline{(L+h+i\alpha)^{-1}(-i\alpha)}\right)$$
and then:
\begin{eqnarray}
0 & = & 
\Re\left(\overline {(L+h+i\alpha)^{-1}(-i\alpha)}\right)\nonumber\\
& = & \Re\left(-{i \overline{\alpha}\over h}  + 
{1\over h} \overline{(-i\alpha)
(L+h+i\alpha)^{-1}(-i\alpha)}\right)\nonumber\\
& = & \Re\left({1\over h}
\overline{\alpha(L+h+i\alpha)^{-1}\alpha}\right) .
\label{sadle5} 
\end{eqnarray}
where in the first step we used again the Schwinger-Dyson identity.
Now applying the inequality (\ref{ineq4'}) to (\ref{sadle5}) we get:
$$0 \leq \overline{\alpha(L+h-i\alpha)^{-1}(L+h+i\alpha)^{-1}\alpha}
\leq\Re\left({1\over h}\overline{\alpha_i(L+h+i\alpha)^{-1}\alpha}\right)$$
\begin{equation}
\overline{\alpha(L+h-i\alpha)^{-1}(L+h+i\alpha)^{-1}\alpha}=0
\label{sadle''} 
\end{equation}
Let us consider the inequality (\ref{ineq5'}). We have:
$$ \overline{\alpha^2} \leq f(\alpha)
\left(\overline{
\alpha(L+h-i\alpha)^{-1}(L+h+i\alpha)^{-1}\alpha}\right)^{1/2}=0~.
$$
Notice that $f(\alpha)$ is bounded if
$\overline{\alpha^4}$ is bounded. So we proved that 
the condition (\ref{sadle2}) for the variables $\alpha_i$ must be satisfied,
if the quantity (\ref{sadle1}) is bounded when $\beta \rightarrow\infty$.
This proof also holds when $h\rightarrow 0$. Indeed in 
this case  the stationary condition (\ref{sadle3}) can be expressed by
(\ref{sadle''}) and the inequality (\ref{ineq5'}) holds.

Now in order to complete our proof we must verify that also for a generic
$V_{\lambda}$ (\ref{sadle1})
is bounded at the stationary point when $\beta\rightarrow\infty$.
In this way we will show that (\ref{sadle2}) is not only a necessary 
condition but also a sufficient one. If we evaluate (\ref{sadle1}) with the 
constraint (\ref{sadle2}) we get:
\begin{eqnarray}
& \ & i \left(-{n\over 2} \overline{\Tr}_{V_{\lambda}} (L+h)^{-1} 
+{ \beta\over 2} \left(\mu(V_{\lambda}) - 
h^2 \overline{(L+h)^{-1} \chi(V_\lambda)
(L+h)^{-1}}\right)\right)\nonumber\\
& = & i \left(-{n\over 2} \overline{\Tr}_{V_{\lambda}} (L+h)^{-1} 
+{ \beta\over 2} \left(\mu(V_{\lambda}) - 
\overline{\chi_{V_\lambda}} \right)\right)\nonumber\\
& = & -i~{n\over 2} \overline{\Tr}_{V_{\lambda}} (L+h)^{-1} 
\nonumber
\end{eqnarray}
where we used again the Schwinger-Dyson identity. For infrared finite graphs 
the last expression is always bounded (also when $h\rightarrow 0$), 
see (\ref{limtr}). This completes our proof. $\Box$.


\begin{thebibliography}{99}

\bibitem{forte} Burioni, R. and Cassi, D.: Geometrical universality in 
vibrational dynamics. Mod. Phys. Lett. B {\bf 11}, 1095-1101, (1997)
\bibitem{debole} Burioni, R. and Cassi, D.: Universal properties of spectral
dimension. Phys. Rev. Lett. {\bf 76} 1091-1093, (1996)
\bibitem{invmwg} Burioni, R., Cassi, D. and Vezzani, A.: in preparation
\bibitem{mwg} Cassi, D.: Local vs. average behavior  on inhomogeneous structures:
recurrence on the average and a further extension of Mermin-Wagner theorem
on graphs. Phys. Rev. Lett. {\bf 76} 2941-2944, (1996) 
\bibitem{FSS}
Fr\"ohlich, J., Simon, B. and Spencer, T.: Infrared bound, phase transitions and
continuous symmetry breaking. Commun. Math. Phys., {\bf 50}, 79-85 (1976)
\bibitem{gant} Gantmacher, F.R.: The Theory of Matrices. New York: Chelsea Publishing
Company 1959  
\bibitem{hhw}
Hattori, K., Hattori, T. and Watanabe, H.: Gaussian Field Theories 
on General Networks
and the Spectral Dimension. Progr. of Theor. Phys. Suppl. {\bf 92}, 108-143 
(1987) 
\bibitem{mw} Mermin, N.D.: Absence of ordering in certain classical systems.
Journ. of Math. Phys. {\bf 8}, 1061-1064 (1967)
\bibitem{gerl} Mohar, B.: The spectrum of an infinite graph. Linear Algebra Appl. {\bf 48},
245-256 (1982)
\bibitem{woess} Mohar, B. and Woess, W.: A survey on spectra of infinite graphs. Bull. London 
Math. Soc. {\bf 21}, 209-234 (1989)
\end{thebibliography}
\end{document}